\documentclass[11pt]{aims}   	
\usepackage{amsmath}
  \usepackage{paralist}
  \usepackage{graphics} 
  \usepackage{epsfig} 
\usepackage{graphicx}  \usepackage{epstopdf}
\usepackage{subfigure}
 \usepackage[colorlinks=true]{hyperref}
\hypersetup{urlcolor=blue, citecolor=red}

\def\currentvolume{X}
 \def\currentissue{X}
  \def\currentyear{200X}
   \def\currentmonth{XX}
    \def\ppages{X--XX}
     \def\DOI{10.3934/xx.xx.xx.xx}

\newtheorem{theorem}{Theorem}

\theoremstyle{definition}

\title[HIV dynamics with Latent Infection \& Mutation] 
      {An in-host model of HIV incorporating latent infection and viral mutation}

\author[Stephen Pankavich and Deborah Shutt]{}

\subjclass{Primary: 37N25, 92B05; Secondary: 34D20.}
 \keywords{HIV, in-host dynamics, latent infection, viral mutation, ART}

 \email{pankavic@mines.edu}
 \email{dshutt@mines.edu}

\thanks{This work is supported by National Science Foundation grant DMS-1211667}

\begin{document}
\maketitle

\centerline{\scshape Stephen Pankavich }
\medskip
{\footnotesize
   \centerline{Department of Applied Mathematics and Statistics}
    \centerline{Colorado School of Mines}
   \centerline{ Golden, CO 80401, USA}
} 

\medskip

\centerline{\scshape Deborah Shutt}
\medskip
{\footnotesize
   \centerline{Department of Applied Mathematics and Statistics}
    \centerline{Colorado School of Mines}
   \centerline{ Golden, CO 80401, USA}
}

\bigskip

 \centerline{(Communicated by the associate editor name)}

\begin{abstract}
We construct a seven-component model of the in-host dynamics of the Human Immunodeficiency Virus Type-1 (i.e, HIV) that accounts for latent infection and the propensity of viral mutation.  A dynamical analysis is conducted and a theorem is presented which characterizes the long time behavior of the model.  Finally, we study the effects of an antiretroviral drug and treatment implications.
\end{abstract}

\section{Introduction and Model}
The human immune system is a complex network of interacting cells, cell products, and cell-forming tissues that protects the body from pathogens, destroys infected and malignant cells, and removes cellular debris \cite{biomath}.  A key component of this system is the white blood cell known as a T cell.  
In particular, the CD4$^+$ T cell moves throughout the body, identifying bacteria and viruses and directing the immune system's attack.  
Within healthy individuals the blood concentration of these cells is relatively constant, around $10^6$ cells/ml \cite{AlanNP}.
%
%
HIV molecularly recognizes CD4$^+$T cells as compatible for viral replication.  As HIV  is a retrovirus, it replicates within a host via reverse transcription using its RNA and an enzyme (reverse transcriptase) to create a strand of HIV DNA, called a provirus, which carries its genetic information.  Once the provirus is created within a CD4$^+$T cell, its immunological function ceases and it is used to create new virions \cite{NM}.  
Interestingly, reverse transcription is quite error prone, and the probability of mutation is high (around 22$\%$ of infected cells carry proviral genomes with at least one mutation \cite{AlanRFP}).  
Without being activated by antigens, an infected CD4$^+$T cell, containing a provirus, fails to produce new virus particles.  In such a case, the T cell is referred to as latently infected \cite{Latent, RP, RP2}.



\begin{figure}[!t]
   \centering 
   \subfigure[Healthy T cells]{\includegraphics[width=.35\textwidth]{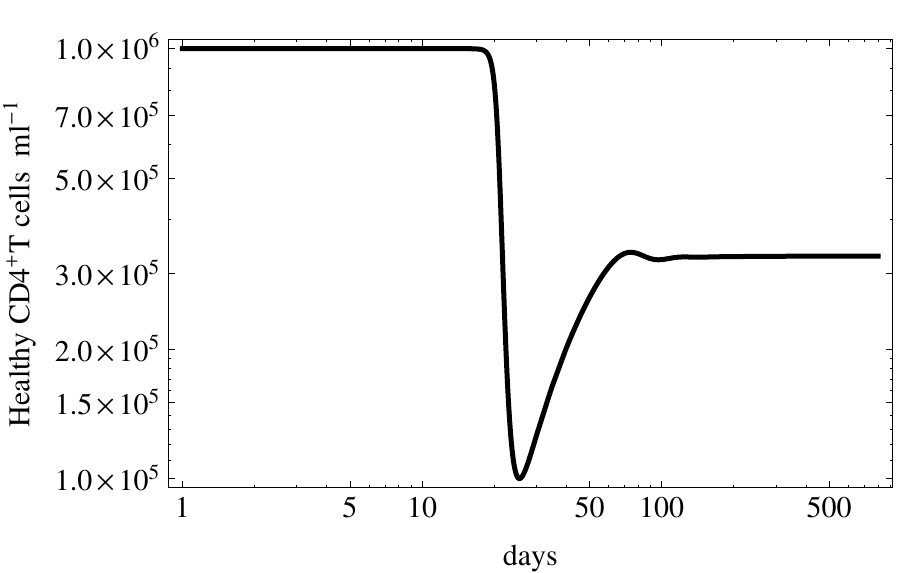}} \hspace{0.5in}
   \subfigure[Wild-type HIV Strain]{\includegraphics[width=.35\textwidth]{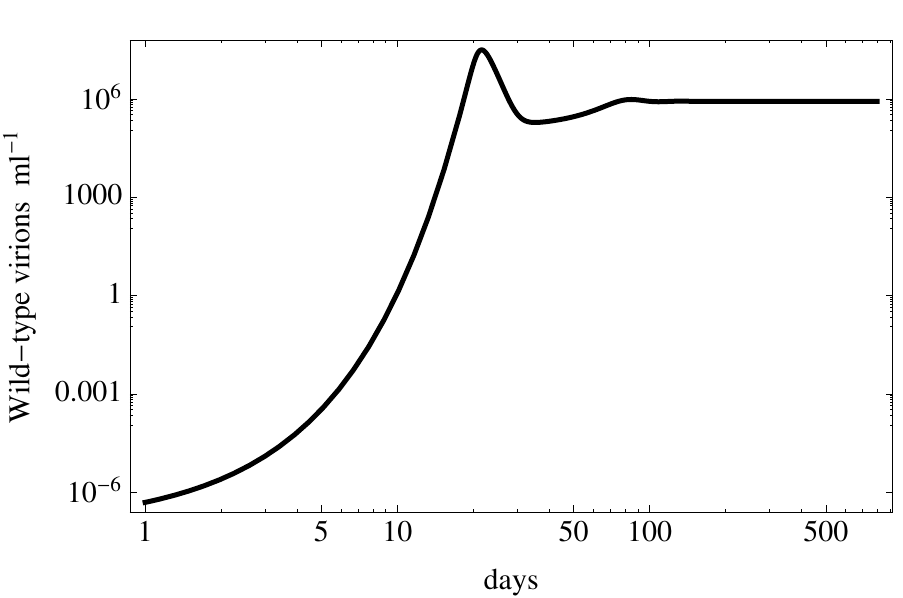}}
   \subfigure[Mutated HIV Strain]{\includegraphics[width=.35\textwidth]{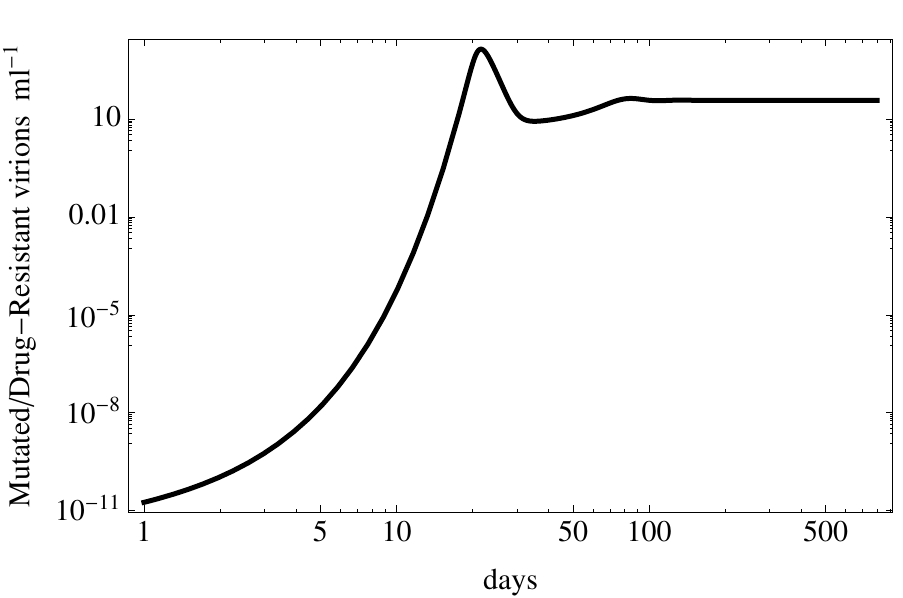}} \hspace{0.5in}
   \subfigure[Total Virus Load]{\includegraphics[width=.35\textwidth]{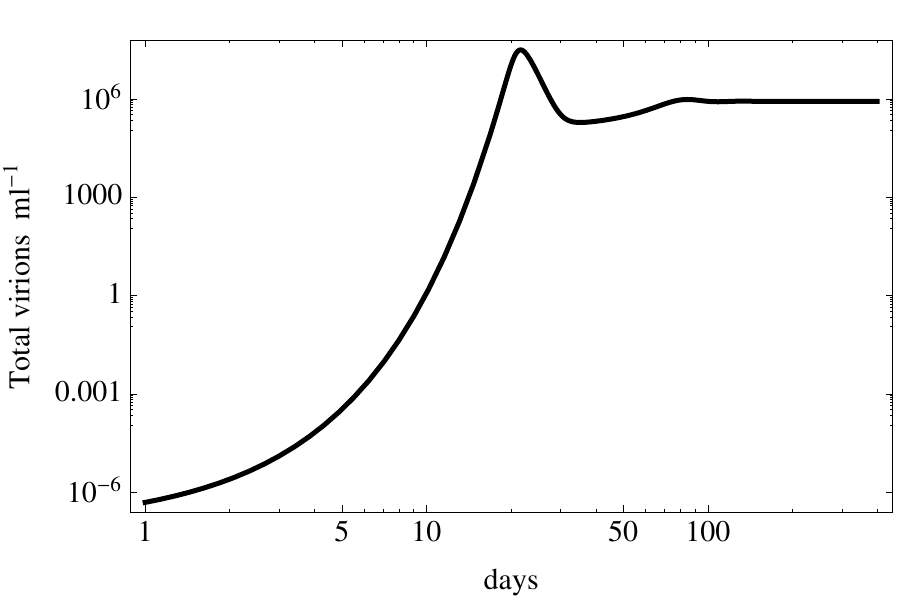}}
   \caption{\footnotesize A log-log simulation of \eqref{7CM} with parameters given in Table \ref{param}.
Notice that the wild-type viral strain is dominant.}
\label{simulation}
\vspace{-0.1in}
\end{figure}

The following model describes interactions between T cells and the populations of a wild-type and mutation of HIV.  It incorporates latent infection of T cells and accounts for a single mutation within the virus.  Specifically, the model is comprised of seven nonlinear ODEs given by
\begin{equation}
\label{7CM}
\tag{7CM}
\begin{small}
\left.
\begin{split}
\frac{dT}{dt}&=\lambda-d_T T - k_s T V_s - k_r T V_r\\
\frac{dI_s}{dt}&=(1-p)(1-\mu) k_s T V_s + \alpha_s L_s - d_I I_s\\
\frac{dL_s}{dt}&=p(1-\mu) k_s T V_s - \alpha_s L_s - d_L L_s\\
\frac{dV_s}{dt}&=N_s d_I I_s - d_V V_s\\
\frac{dI_r}{dt}&=(1-p)\mu k_s T V_s + (1-p) k_r T V_r + \alpha_r L_r - d_I I_r\\
\frac{dL_r}{dt}&=p \mu k_s T V_s +p k_r T V_r - \alpha_r L_r - d_L L_r\\
\frac{dV_r}{dt}&=N_r d_I I_r - d_V V_r.
\end{split}
\right \}
\end{small}
\end{equation}
\noindent Here $T$ denotes the population of healthy T cells whose maturity rate is $\lambda$ and death rate is $d_T$.  The parameter $k_s$ is the rate of infection by the wild-type virus, whose population is denoted by $V_s$.  Note that the T cells infected by the original strain, $k_s T V_s$, contribute to several other populations. $I_s$ and $L_s$ denote the populations of wild-type actively or latently infected T cells, respectively.  The proportion of T cells which become latently infected is given by $p$.   $I_r$ and $L_r$ are the populations of T cells that are actively and latently infected by the mutated strain, while the mutation rate of the wild-type virus is $\mu \in [0,1]$.  Note that T cells which are infected by the wild-type strain may create a mutated provirus and thus the terms $(1-p) \mu k_s T V_s$ and $p\mu k_s T V_s$ are included.
The activation rates of latently infected cells to actively infected are $\alpha_s$ and $\alpha_r$ for the wild-type and mutated strains.  The death rates of actively and latently infected T cells, regardless of strain, are $d_I$ and $d_L$. $N_s$ and $N_r$ are the numbers of virions (wild and mutated) released by an infected T cell over its lifespan.  Finally, $d_V$ is the viral clearance rate for both strains.  See the review \cite{AlanRuy} for more information. Parameter values are in Table \ref{param}, while Fig. \ref{simulation} presents a simulation of \eqref{7CM} with this data.  Both strains persist in the simulation, but the wild-type dominates the mutated strain by five orders of magnitude. 
In subsequent sections we will utilize analytic and computational methods to study \eqref{7CM} and discuss the effects of antiretroviral drugs on these populations.

\begin{table}[!t]
\centering
\caption{Variables and Parameters}
\label{param}
\begin{footnotesize}
\begin{tabular}{p{.5cm}  p{7.5cm} p{2.75cm} p{0.9cm}}
\hline
& Variables/Parameters & (Initial) Value & \hspace{-.5cm}Reference\\
\hline
\multicolumn{4}{l}{}\\
$T$ & Uninfected CD4$^+$ T cells & $10^6~\mathrm{ml}^{-1}$ & \cite{AlanNP}\\
$I_s$ & CD4$^+$ T cells actively infected by wild-type virions & 0 \\
$L_s$ & CD4$^+$ T cells latently infected by wild-type virions  & 0 \\
$V_s$ & Wild-type HIV virions & $10^{-6}~\mathrm{ml}^{-1}$ & \cite{AlanDenise}\\
$I_r$ & CD4$^+$ T cells actively infected by mutated virions & 0 \\
$L_r$ & CD4$^+$ T cells latently infected by mutated virions  & 0 \\
$V_r$ & Mutated/Drug-resistant HIV population size & 0\\
\\
$\lambda$ & Rate of supply of  immunocompetent CD4$^+$ T cells &$10^4~\mathrm{ml}^{-1}~\mathrm{day}^{-1}$ & \cite{AlanRFP}\\
$d_T$ & Death rate of uninfected T cells & $0.01~\mathrm{day}^{-1}$& \cite{AlanRFP} \\
$k_s$ & Infection rate of T cells by wild-type HIV &  $2.4\times10^{-8}~\mathrm{ml}~\mathrm{day}^{-1}$ & \cite{AlanRFP}\\
$k_r$ & Infection rate of T cells by mutated HIV  &$2.0\times10^{-8}~\mathrm{ml}~\mathrm{day}^{-1}$ & \cite{AlanRFP}\\
$p$ & Rate of latent infection in T cell population & $0.1$ & \cite{Latent}\\
$d_I$ & Death rate of actively infected T cells & $1~\mathrm{day}^{-1}$ & \cite{AlanRFP}\\
$d_L$ & Death rate of latently infected T cells &$4\times10^{-3}~\mathrm{day}^{-1}$ & \cite{Latent}\\
$\alpha_s$ & Activation rate for latently infected wild-type T cells  & $0.01$ & \cite{Latent}\\
$\alpha_r$ & Activation rate for latently infected mutated T cells & $0.01$  & \cite{Latent}\\
$N_s$ & Burst size of wild-type virus & 3000 & \cite{AlanRFP}\\
$N_r$ & Burst size of mutated/drug-resistant virus & 2000 & \cite{AlanRFP}\\
$\mu$ & Mutation rate from wild-type to mutated & $3\times10^{-5}$ & \cite{AlanRFP}\\
$d_V$ & Clearance rate of free virus & $23~\mathrm{day}^{-1}$ & \cite{AlanRFP}\\
\hline
\end{tabular}
\end{footnotesize}
\vspace{-0.1in}
\end{table}
%

\section{Analysis and Dynamics}
\begin{figure}[!t]
   \centering 
   \includegraphics[height=.45\textwidth]{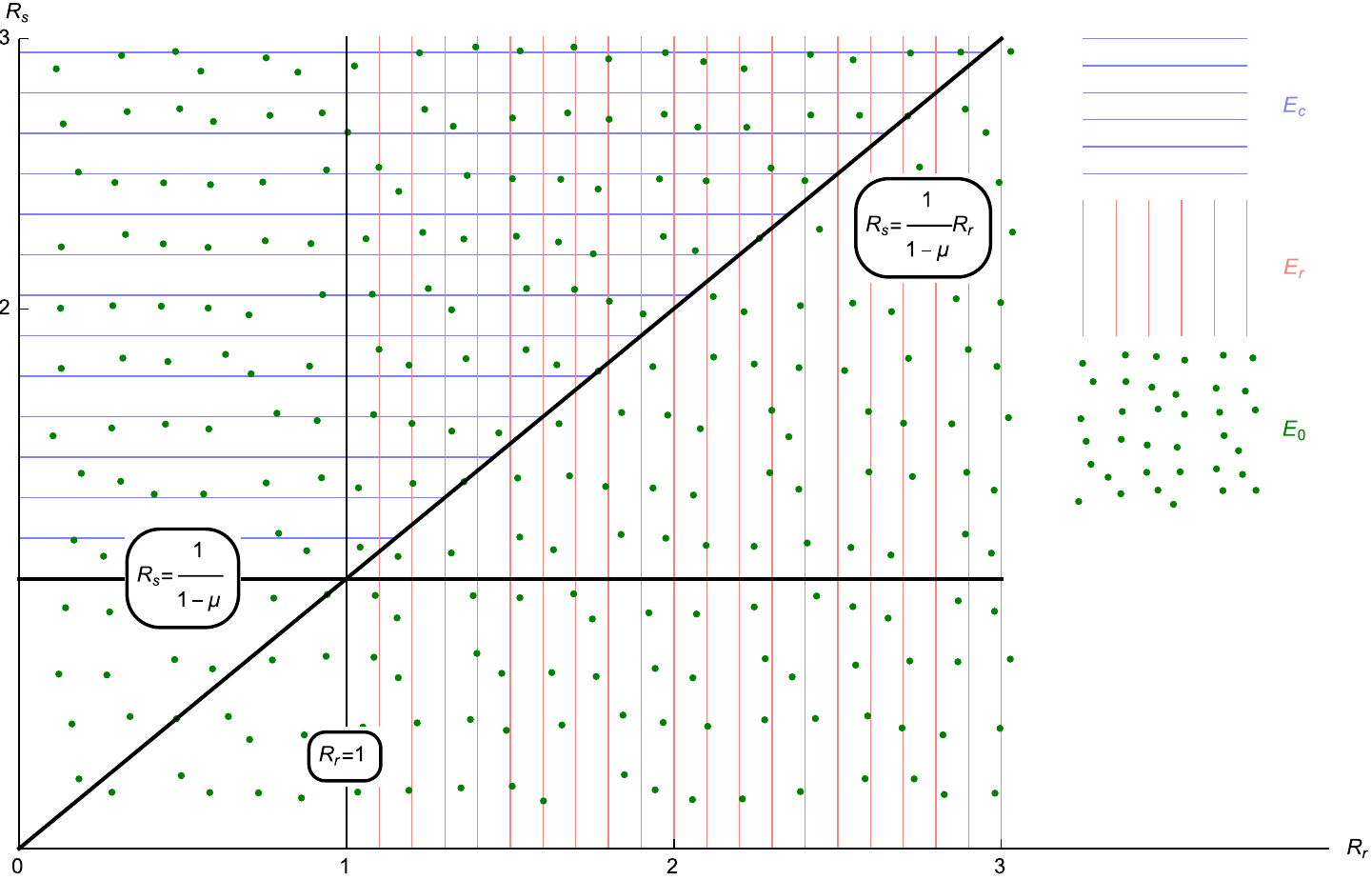}
   \caption{\footnotesize Regions of biologically-relevant existence for steady states in the ($R_r$,$R_s$)-plane.  
   }
   \label{existence}
\end{figure}

With the model finalized, we turn to an analysis of its behavior. First, well-posedness of solutions to \eqref{7CM} can be shown using the Picard-Lindel$\ddot{\mbox{o}}$f Theorem and known \emph{a priori} bounds.  The details are standard so we omit them.  Instead, we focus on the asymptotic behavior of \eqref{7CM} as $t\to\infty$.
A number of long calculations show that exactly three steady states exist - $E_0$, $E_r$ and $E_c$, given by
\begin{equation*}
\begin{scriptsize}
\centering
\mathbf{E_0} = \left(
\def\arraystretch{2}\begin{array}{c}
\frac{\lambda}{d_T}\\
0\\ 0\\ 0\\ 0\\ 0\\ 0
\end{array}\right)
 \
\mathbf{E_r} = \left(
\def\arraystretch{2}\begin{array}{c}
\frac{\lambda}{d_T R_r}\\
0\\ 0\\ 0\\
\frac{d_T d_V}{k_r N_r d_I}(R_r -1)\\
\frac{\lambda p}{R_r (d_L + \alpha_r)}(R_r - 1)\\
\frac{d_T}{k_r}(R_r - 1)
\end{array}\right)
 \
\mathbf{E_c} = \left(
\def\arraystretch{2}\begin{array}{c}
\frac{\lambda}{d_T R_c}\\
\frac{d_T d_V (\sigma_c - 1)}{d_I k_s N_s (1-\mu)(\sigma - 1)}(R_c - 1)\\
\frac{\lambda p (\sigma_c - 1)}{R_s(d_L + \alpha_s)(1 - \mu)(\sigma - 1)}(R_c - 1)\\
\frac{d_T(\sigma_c-1)}{k_s(1-\mu)(\sigma - 1)}(R_c - 1)\\
\frac{\mu d_T d_V}{d_I k_r N_r (1-\mu)(\sigma - 1)}(R_c - 1)\\
\frac{\mu \lambda p}{R_r (d_L + \alpha_r)(1-\mu)(\sigma - 1)}(R_c - 1)\\
\frac{\mu d_T}{k_r (1-\mu)(\sigma - 1)}(R_c - 1)\\
\end{array}\right)
\end{scriptsize}
\end{equation*} 
where the newly-introduced parameters $R_r$, $R_s$, $R_c$, $\sigma$ and $\sigma_c$ are defined as
$$\begin{gathered}
R_r = \frac{\lambda k_r N_r (\alpha_r + (1-p)d_L)}{d_T d_V (d_L + \alpha_r)}, \quad R_s = \frac{\lambda k_s N_s (\alpha_s + (1-p)d_L)}{d_T d_V (d_L + \alpha_s)}\\
R_c = (1 - \mu) R_s, \quad \sigma = \frac{R_s}{R_r}, \quad \sigma_c = (1-\mu) \sigma.
\end{gathered}
$$
%
%
$E_0$ is the infection-free steady state, in which all populations but $T$ are zero.  Within $E_r$, the mutation dominant steady state, only the mutated strain remains present as $I_s$, $L_s$ and $V_s$ are zero. $E_c$ is the state in which both strains persist, and hence is referred to as the coexistence steady state. 
To represent biological quantities, all steady state components must be nonnegative. Clearly, this is the case for $E_0$.  Since all parameters are positive, we see that $R_r>1$ implies that $E_r$ satisfies this requirement.  Similarly, $E_c$ requires $R_c>1$ and $\sigma_c>1$, which implies $R_s>\frac{1}{(1-\mu)}$ and $R_s>\frac{1}{(1-\mu)}R_r$.  Note that $\sigma_c>1$ implies $\sigma>1$. Fig. \ref{existence} displays the regions in the $(R_r, R_s)$-plane for which steady states are nonnegative.

Finally, we wish to determine whether such states correspond to locally asymptotically stable (l.a.s) or unstable equilibria.
In this vein, we prove

\begin{theorem}
\label{T1}
The stability properties of the equilibria depend only upon the two parameters $R_r$ and $R_s$.  Moreover, 
\begin{enumerate}
\item The infection-free steady state $E_0$ is l.a.s. if and only if 
$$R_s < 1/(1-\mu) \ \ \mathrm{and} \ \ R_r < 1,$$ and it is unstable otherwise.
\vspace{0.05in}
\item The steady state with only the mutated virus, $E_r$  is l.a.s. if and only if 
$$R_s < R_r/(1-\mu)\ \ \mathrm{and} \ \  R_r > 1,$$ and it is unstable otherwise.
\vspace{0.05in}
\item The coexistence steady state $E_c$ is l.a.s. if and only if 
$$R_s > 1/(1 - \mu) \ \ \mathrm{and}  \ \ R_s > R_r/(1-\mu),$$ and it is unstable otherwise. 
\end{enumerate}
\end{theorem}

\begin{proof}
As in \cite{CP, Peter} we consider the linearization of the system to justify conclusions about the its local asymptotic behavior.
To begin, we consider the state $E_0$, and evaluate the Jacobian of the system at this point. In particular, we are interested in the eigenvalues of the matrix
%
\begin{equation*}
\begin{scriptsize}
\centering
\mathbf{\nabla\bold{f}(E_0)} =\left(
\def\arraystretch{2}\begin{array}{ccccccc}
 -d_T & 0 & 0 & -\frac{\lambda  k_s}{d_T} & 0 & 0 &
   -\frac{\lambda  k_r}{d_T} \\
 0 & -d_I & \alpha _s & \frac{(p-1) \lambda  (\mu -1)
   k_s}{d_T} & 0 & 0 & 0 \\
 0 & 0 & -d_L-\alpha _s & -\frac{p \lambda  (\mu -1) k_s}{d_T}
   & 0 & 0 & 0 \\
 0 & d_I N_s & 0 & -d_V & 0 & 0 & 0 \\
 0 & 0 & 0 & -\frac{(p-1) \lambda  \mu  k_s}{d_T} &
   -d_I & \alpha _r & -\frac{(p-1) \lambda  k_r}{d_T}
   \\
 0 & 0 & 0 & \frac{p \lambda  \mu  k_s}{d_T} & 0 & -d_L-\alpha
   _r & \frac{p \lambda  k_r}{d_T} \\
 0 & 0 & 0 & 0 & d_I N_r & 0 & -d_V.
\end{array}
\right)_.
\end{scriptsize}
\end{equation*}
The seven eigenvalues are  $\lambda_1 = -d_T$; 
 $\lambda_{2,3,4} $ defined by the roots of the cubic $$a_3 x^3 +a_2 x^2 + a_1 x +a_0$$  where $a_3 = 1$ and  $a_2 = d_I d_T + d_L d_T + \alpha_r d_T + d_T d_V$ are both positive, and 
 $$\begin{gathered}
 a_1 = {d_T}^{2}(d_V + d_I)(d_L + \alpha_r)+ {d_T}^{2} d_V d_I - d_T d_I (1-p)\lambda k_r N_r,\\
 a_0 = {d_T}^{3} d_V d_I (d_L + \alpha_r) - d_I {d_T}^{2}\lambda k_r N_r (\alpha_r + (1-p)d_L)
 \end{gathered}$$
and $\lambda_{5,6,7}$ defined as the roots of the cubic polynomial $$b_3 x^3 +b_2 x^2 + b_1 x +b_0$$  where  $b_3 = 1$ and $b_2 = d_I d_T + d_L d_T + \alpha_s d_T + d_T d_V$ are both positive, and 
$$\begin{gathered}
b_1 = {d_T}^{2} (d_V + d_I)(d_L + \alpha_s)+ {d_T}^{2} d_V d_I - d_T d_I(1 - \mu)(1-p)\lambda k_s N_s),\\
b_0 = {d_T}^{3} d_V  d_I (d_L + \alpha_s) - {d_T}^{2} d_I (1 - \mu) \lambda k_s N_s (\alpha_s + (1-p) d_L).
\end{gathered}$$
%
\begin{figure}[!t]
   \centering 
   \subfigure[Behavior in Region 1]{\includegraphics[width=.45\textwidth]{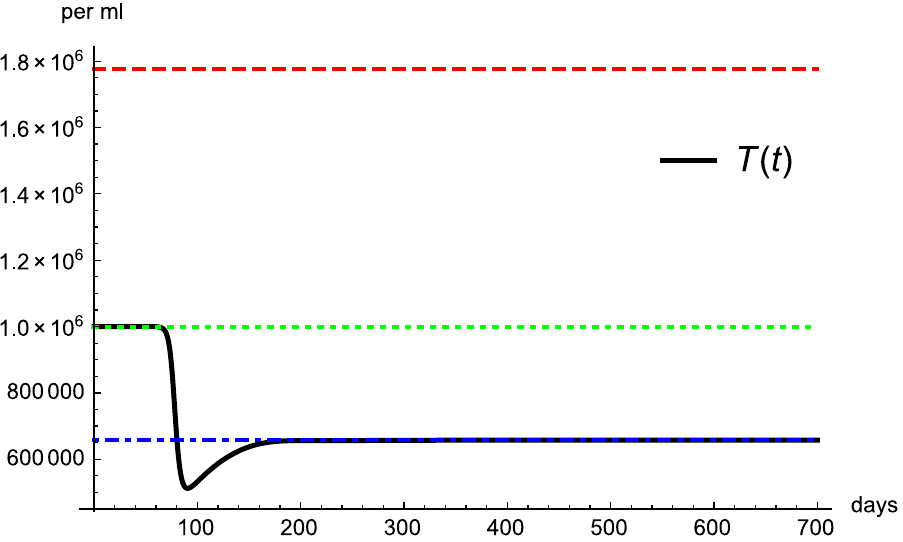}}
   \subfigure[Behavior in Region 2]{\includegraphics[width=.45\textwidth]{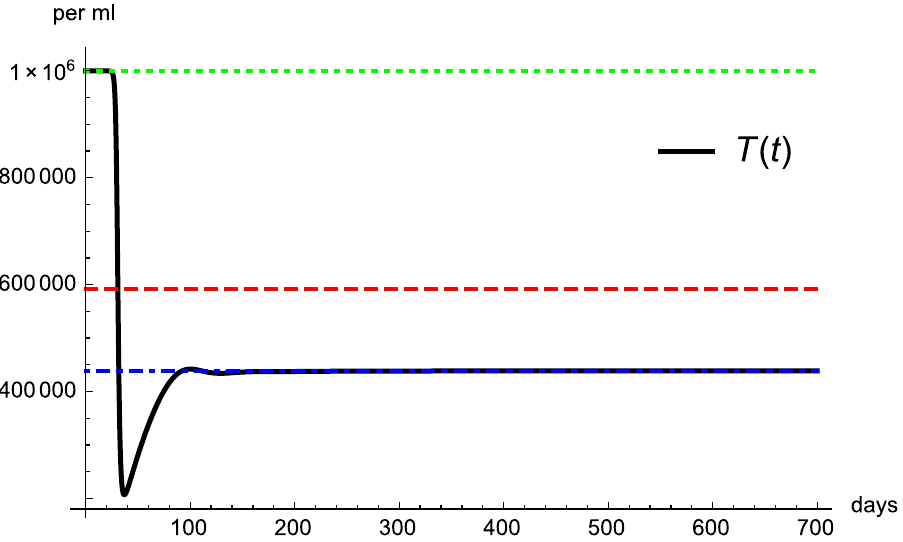}}
   \subfigure[Behavior in Region 3]{\includegraphics[width=.45\textwidth]{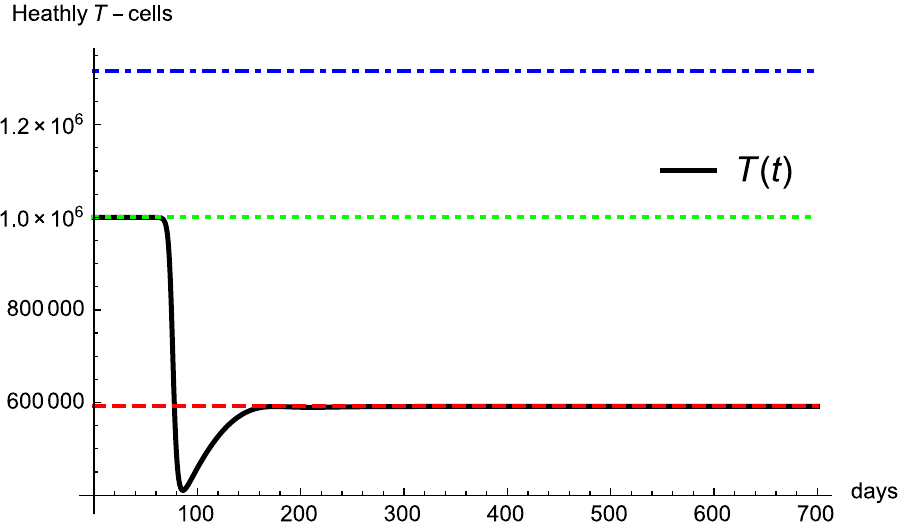}}
   \caption{\footnotesize Stability of steady states in various regions of the $(R_r,R_s)$-plane.  
   In each figure, the $T$ component of $E_0$, $E_r$, and $E_c$ is represented by dotted, dashed, and dot-dashed
   lines, respectively, while the solid curve is the solution $T(t)$ with $T(0) = 10^{6}$. }
   \label{Eqs}
   \vspace{-0.1in}
\end{figure}

\noindent It follows from the Hartman-Grobman Theorem that $E_0$ is l.a.s if and only if $Re(\lambda_i)<0$ for all eigenvalues, $\lambda_i$, of $\nabla\bold{f}(E_0)$.  By the Routh-Hurwitz Theorem, the roots of the cubics possess negative real parts if and only if $a_0, a_1, a_2, a_3 > 0$ and $a_2 a_1 >a_3 a_0$.  Clearly, $a_2, a_3, b_2, b_3 > 0$.  For the first cubic, $a_1, a_0 > 0$ and $a_2 a_1 > a_0$ are satisfied for $R_r < 1$.  For the second cubic, the corresponding requirements are satisfied if and only if $R_s < \frac{1}{1-\mu}$.  Hence, $E_0$ is l.a.s for the specified conditions. 
A similar analysis proves the second and third conclusions of the theorem.  For brevity, we omit the details but take a computational approach to demonstrate their validity.

Fig. \ref{existence} displays the regions in which nonnegative steady states exist.   In the area of the $(R_r,R_s)$-plane where $R_s>\frac{1}{1-\mu}$ and $R_r<1$ (Region 1), both $E_0$ and $E_c$ exist.  For $R_s>\frac{1}{1-\mu}R_r$ and $R_r>1$ (Region 2), all three steady states exist.  For $R_s<\frac{1}{1-\mu}R_r$ and $R_r>1$ (Region 3), both $E_0$ and $E_r$ exist.
Simulations are used to demonstrate the stability of coexisting states.  We first begin with parameter values placing $(R_r, R_s)$ in Region 1. Note in Fig. \ref{Eqs}(a) that  $T(t)$ (solid) tends to the corresponding value of $E_c$ (dot-dashed).  Next, $(R_r, R_s)$ is adjusted to lie within Region 2, in which all steady states exist. Fig. \ref{Eqs}(b) shows that $T(t)$ still tends to the corresponding value in $E_c$.  Thus, simulations validate the results of the theorem, namely that within Regions 1 and 2, $E_c$ is l.a.s. while $E_0$ and $E_r$ are unstable.  Using $(R_r, R_s)$ within Region 3, Fig. \ref{Eqs}(c) demonstrates that $T(t)$ tends to the corresponding value of $E_r$ (dashed) as $t\to\infty$.  Thus in Region 3, $E_r$ is l.a.s.  While only the healthy T cell population is shown, all other populations
display the corresponding long term behavior.  These findings and the theorem are summarized by Fig.~\ref{stability}. 
  
\begin{figure}[!t]
   \centering
   \includegraphics[height=.45\textwidth]{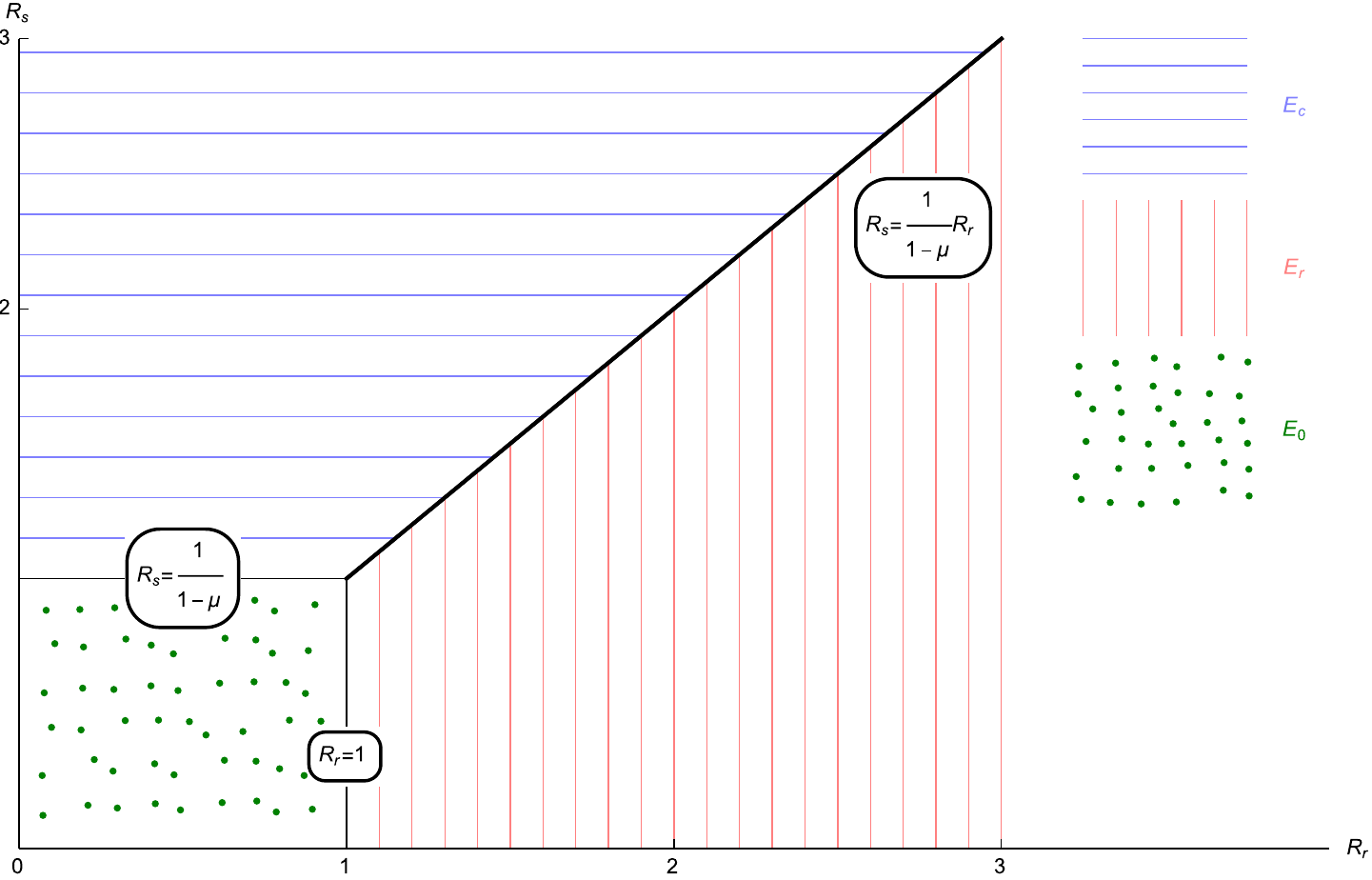}
   \caption{\footnotesize Regions of stability of equilibrium points in the ($R_r$,$R_s$)-plane}
   \label{stability}
   \vspace{-0.1in}
\end{figure}
\end{proof}

\section{Drug Therapy}

In the previous section it was shown that the long term behavior of \eqref{7CM} depends crucially upon parameter values, and even in the presence of viral mutation, the virus may be completely cleared from the body. Contrastingly, the mutated strain may dominate the dynamics, which can lead to drug resistance. Next, we consider the introduction of antiretroviral therapy (ART), specifically reverse transcriptase inhibitors (RTIs). These drugs hinder the replication process and decrease the rate of infection, thus altering $k_s$ and $k_r$.     
Let $\varepsilon_s$ represent the efficacy rate of the RTI on the wild strain.  We assume the RTI combats this strain, while the mutated strain is more resistant to the drug.  
Hence, the efficacy of the RTI on the drug resistant strain is given by $\varepsilon_r = \alpha \varepsilon_s$, where $\alpha\in(0,1)$ represents the resistance level.  Note that smaller values of $\alpha$ correspond to greater levels of resistance.  Including ART within the previous model yields:
\begin{equation}
\label{7CMART}
\tag{7CM$_\varepsilon$}
\begin{scriptsize}
\left.
\begin{split}
\frac{dT}{dt}&=\lambda-d_T T -(1-\varepsilon_s) k_s T V_s -(1-\varepsilon_r)k_r T V_r\\
\frac{dI_s}{dt}&=(1-p)(1-\mu) (1-\varepsilon_s)k_s T V_s + \alpha_s L_s - d_I I_s\\
\frac{dL_s}{dt}&=p(1-\mu) (1-\varepsilon_s)k_s T V_s - \alpha_s L_s - d_L L_s\\
\frac{dV_s}{dt}&=N_s d_I I_s - d_V V_s\\
\frac{dI_r}{dt}&=(1-p)\mu(1-\varepsilon_s) k_s T V_s + (1-p)(1-\varepsilon_r) k_r T V_r + \alpha_r L_r - d_I I_r\\
\frac{dL_r}{dt}&=p \mu (1-\varepsilon_s)k_s T V_s +p (1-\varepsilon_r)k_r T V_r - \alpha_r L_r - d_L L_r\\
\frac{dV_r}{dt}&=N_r d_I I_r - d_V V_r.
\end{split}
\right \}
\end{scriptsize}
\end{equation}
Based on similarities between \eqref{7CM} and \eqref{7CMART}, the steady states are merely $E_0, E_r$, and $E_c$ under the transformations $k_s \mapsto (1-\varepsilon_s)k_s$ and $k_r \mapsto (1-\varepsilon_r)k_r$. We denote them by $E_0^\varepsilon$, $E_r^\varepsilon$, and $E_c^\varepsilon$, respectively.
%
%
\begin{figure}[!t]
\centering 
\subfigure[$\alpha = 0.5$]{\includegraphics[width=.48\textwidth]{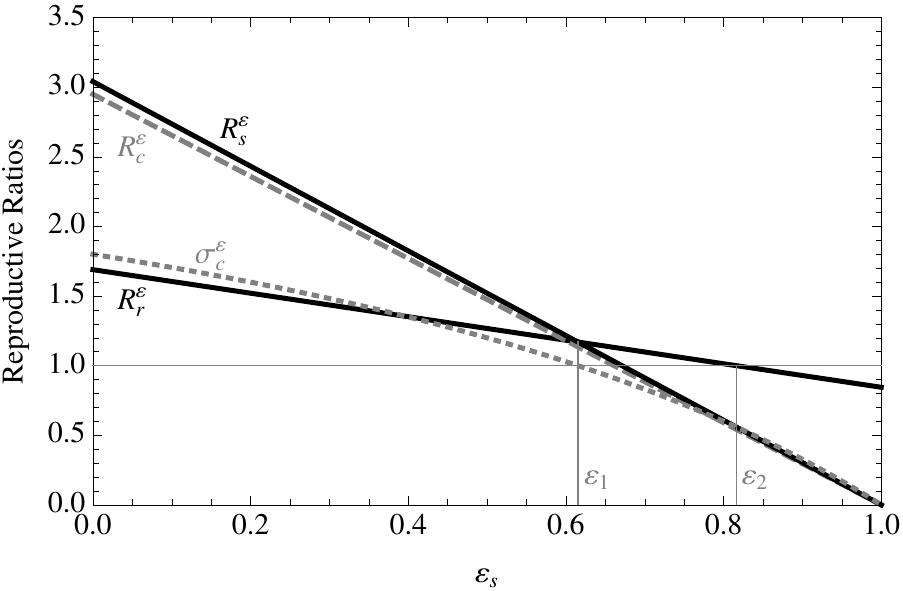}} \quad
\subfigure[$\alpha = 0.2$]{\includegraphics[width=.48\textwidth]{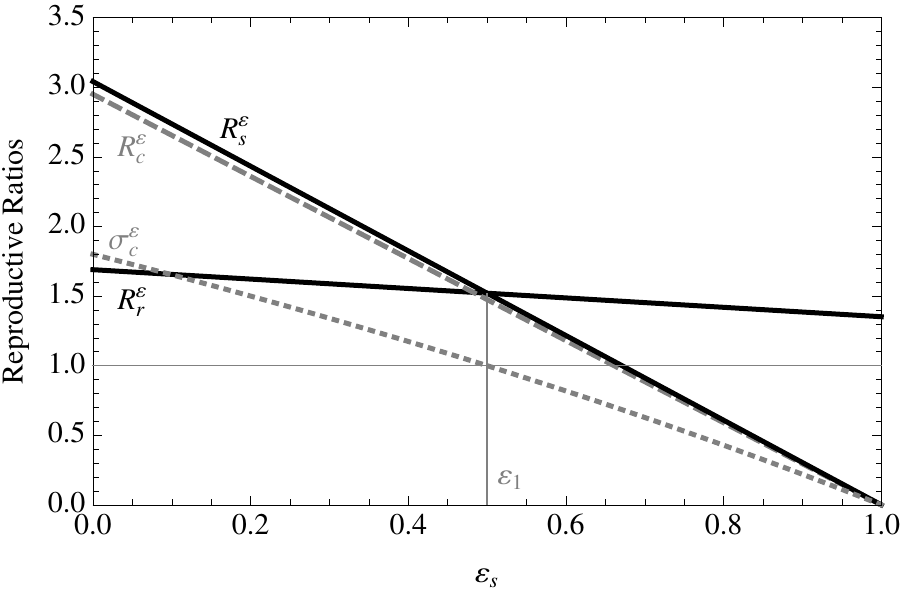}}
\caption{\footnotesize For visual clarification the difference between $R_s^{\varepsilon}$ and $R_c^{\varepsilon}$ has been enlarged by taking $\mu = 3\times{10^{-2}}$ in $R_c^{\varepsilon}$. Note the threshold differences between $\alpha = 0.5$ (left) and  $\alpha = 0.2$ (right).
}
\label{alpha5}
\end{figure}
The new parameters $R_r^{\varepsilon}$ and $R_s^{\varepsilon}$ (as well as $R_c^{\varepsilon}$, $\sigma^{\varepsilon}$ and $\sigma_c^{\varepsilon}$) are defined similarly: 
$$
R_r^{\varepsilon} = (1-\varepsilon_r)R_r , \qquad
R_s^{\varepsilon} = (1-\varepsilon_s)R_s.\\
$$
Hence, the stability properties can be deduced from Theorem \ref{T1} by substituting $R_r = R_r^{\varepsilon}$, and $R_s = R_s^{\varepsilon}$.

We examine how the reproductive ratios are affected by ART for increasing drug efficacy rates assuming a particular resistance level, in this case $\alpha = 0.5$.  In Fig. {\ref{alpha5}}(a), the efficacy rate $\varepsilon_s$ is plotted against reproductive ratios, $R_s^{\varepsilon}$ and $R_r^{\varepsilon}$, along with $R_c^{\varepsilon}$ and $\sigma_c^{\varepsilon}$ to view changes in long term behavior.  The threshold values constituting these changes are denoted by $\varepsilon_1$ and $\varepsilon_2$.
For values of $\varepsilon_s < \varepsilon_1$, we observe $R_c^{\varepsilon} > 1\Leftrightarrow R_s^{\varepsilon} > \frac{1}{1-\mu}$ and $\sigma_c^{\varepsilon} > 1\Leftrightarrow R_s^{\varepsilon} > \frac{1}{1-\mu}R_r^{\varepsilon}$.  Thus by Theorem \ref{T1}, we conclude that solutions tend to $E_c^{\varepsilon}$. Contrastingly, for $\varepsilon_1 < \varepsilon_s < \varepsilon_2$, $R_r^{\varepsilon} > 1$ and $\sigma_c^{\varepsilon} < 1$ thus solutions tend to $E_r^{\varepsilon}$. 
It is not until $\varepsilon_s > \varepsilon_2$, when $R_r^{\varepsilon} < 1$ and $R_c^{\varepsilon} < 1$, that the efficacy is large enough to force viral clearance.

To investigate the change in the long term behavior of the viral loads, we graph the steady state values of $V_r$ and $V_s$ separately against $\varepsilon_s$.  Figs. {\ref{ViralLoad}}(a) and {\ref{ViralLoad}}(b) display the change in the drug resistant virus population. Fig. {\ref{ViralLoad}(a) shows the dramatic increase in the drug resistant population near $\varepsilon_s = \varepsilon_1$, but this scale cannot accurately display values for $\varepsilon_s < \varepsilon_1$.  Thus Fig. {\ref{ViralLoad}}(b), an enlargement, shows the slow increase in this population.  In Fig. {\ref{ViralLoad}}(c), we note the steady decline in the wild population as $\varepsilon_s$ increases and its extinction at $\varepsilon_s = \varepsilon_1$.  Fig. {\ref{ViralLoad}}(d) displays the total virus population.
\begin{figure}[!t]
\centering
  \subfigure[$V_r$ steady state]{\includegraphics[height=.25\textwidth]{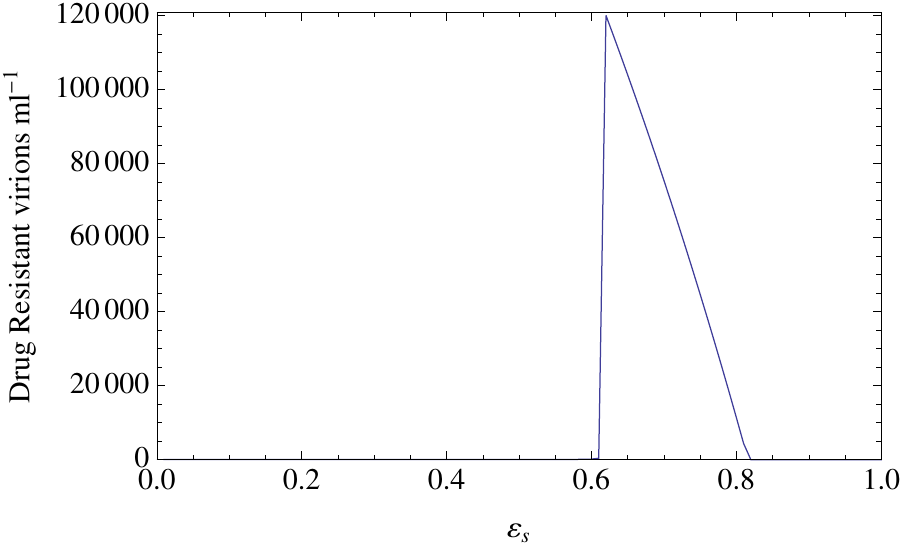}}
  \subfigure[$V_r$ steady state, $\varepsilon_s < \varepsilon_1$]{\includegraphics[height=.25\textwidth]{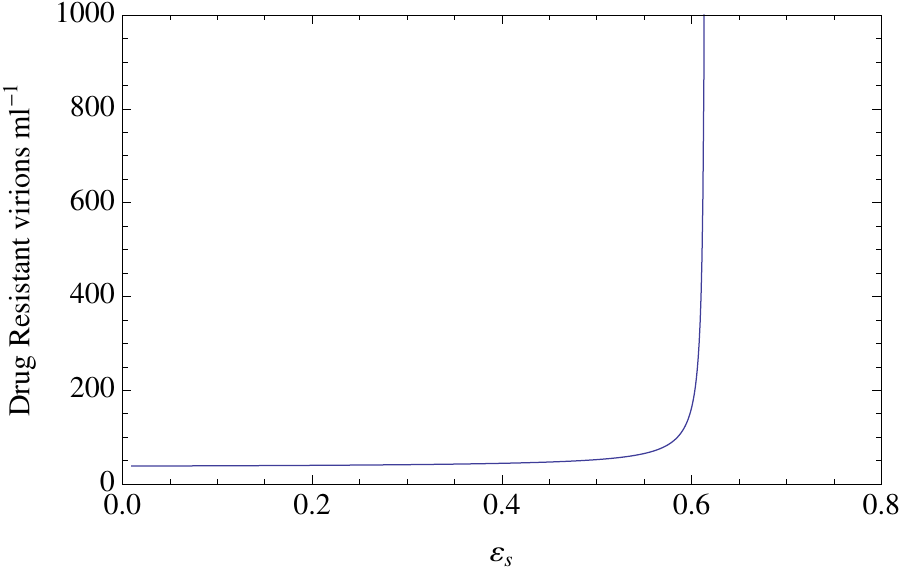}}\\
  \subfigure[$V_s$ steady state]{\includegraphics[height=.25\textwidth]{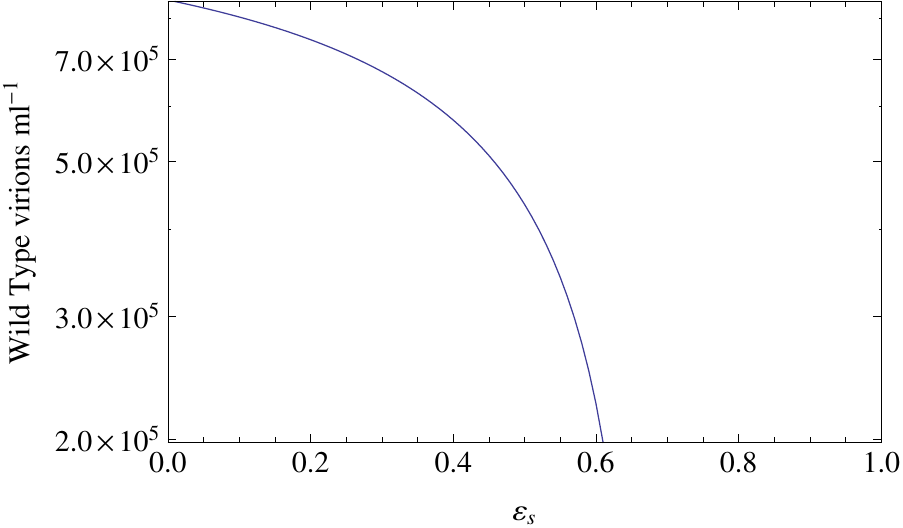}}
  \subfigure[Total virus steady state]{\includegraphics[height=.25\textwidth]{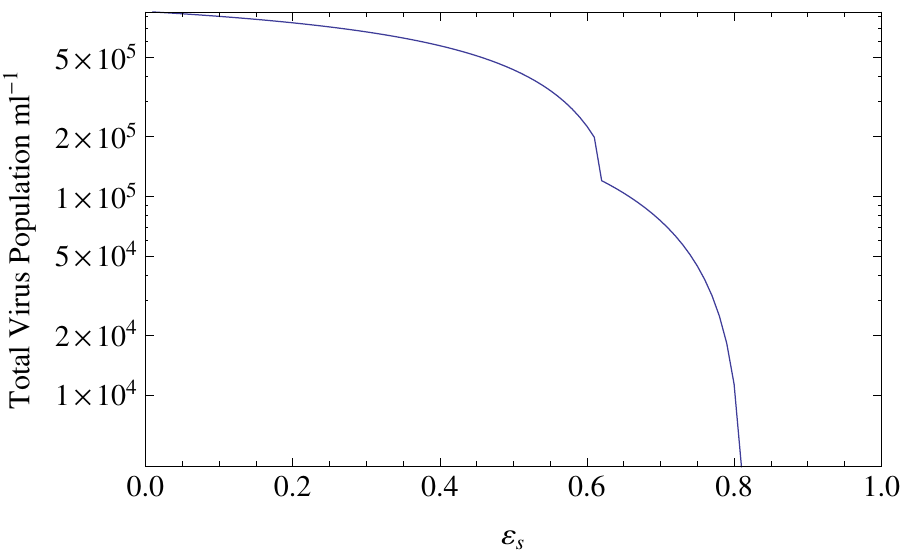}}
  \caption{\footnotesize Steady state values for both wild type and drug resistant virus population with respect to $\varepsilon_s$, assuming $\alpha = 0.5$.}
  \label{ViralLoad}
\end{figure}

The question arises whether a value of $\alpha$ exists which prohibits viral clearance altogether. A higher resistance level of the mutated strain is chosen, and the steady state behavior is examined. Fig. {\ref{alpha5}}(b) shows the reproduction numbers assuming $\alpha = 0.2$. The threshold $\varepsilon_1 = 0.5$ when  $\sigma_c^{\varepsilon} = 1$ indicates the shift from the coexistent state to drug resistant dominance. There is no longer an $\varepsilon_2$ threshold, which implies that no drug efficacy could be large enough to counter the resistance level of the mutant strain.  Thus, the RTI would clear only the wild type virus.
Fig. {\ref{ViralLoad2}}(a) demonstrates that even when $\varepsilon_s = 1$, the resistant steady state remains prohibitive.  In Fig. {\ref{ViralLoad2}}(c), we note the faster decline in the wild population as $\varepsilon_s$ increases and the earlier extinction at $\varepsilon_1$ as compared to Fig. {\ref{ViralLoad}}(c).  Fig. {\ref{ViralLoad2}}(d) displays the total virus population, demonstrating continued viral persistence as $\varepsilon_s \rightarrow 1$, unlike Fig. {\ref{ViralLoad}}(d).
Therefore, viral mutation and drug resistance can completely inhibit the treatment ability of RTIs.


\begin{figure}[!t]
\centering 
\subfigure[$V_r$ steady state]{\includegraphics[height=.25\textwidth]{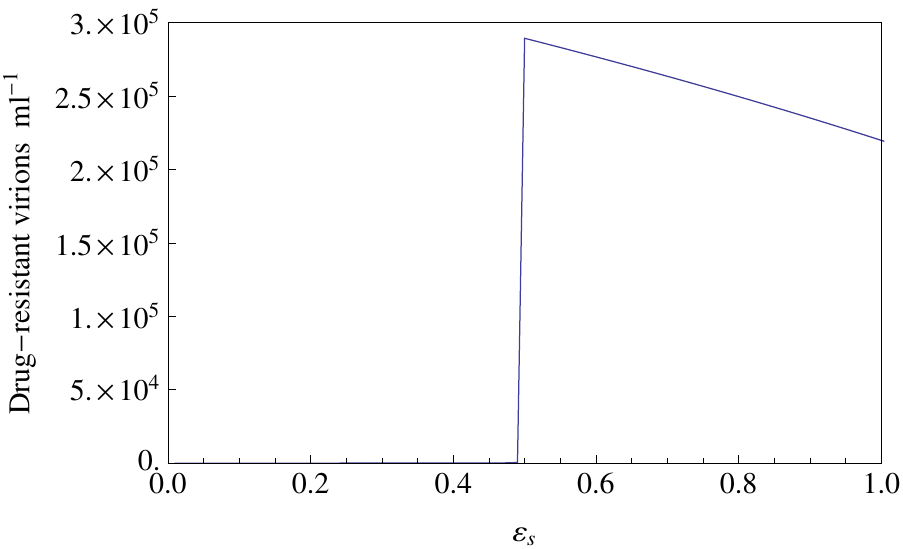}} \qquad 
\subfigure[$V_r$ steady state, $\varepsilon_s < \varepsilon_1$]{\includegraphics[height=.25\textwidth]{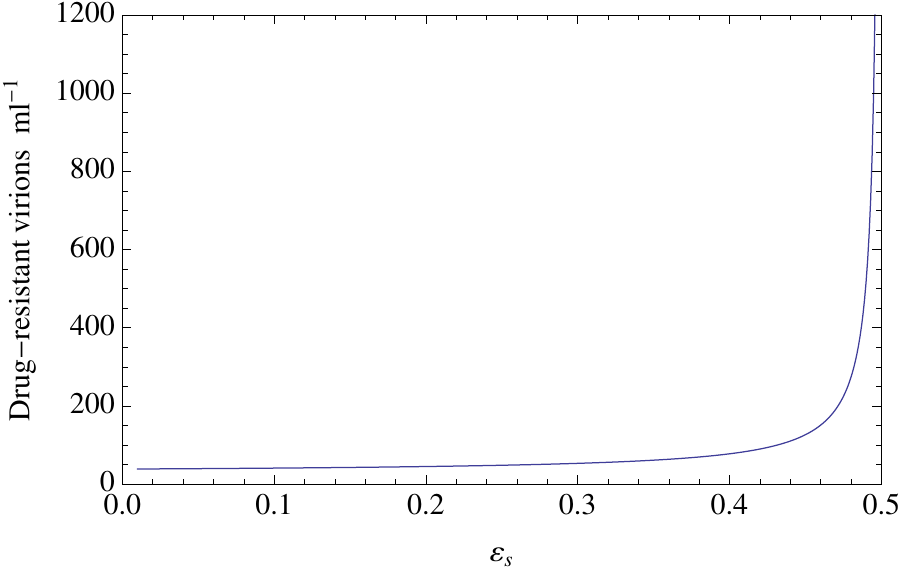}}\\
\subfigure[$V_s$ steady state]{\includegraphics[height=.25\textwidth]{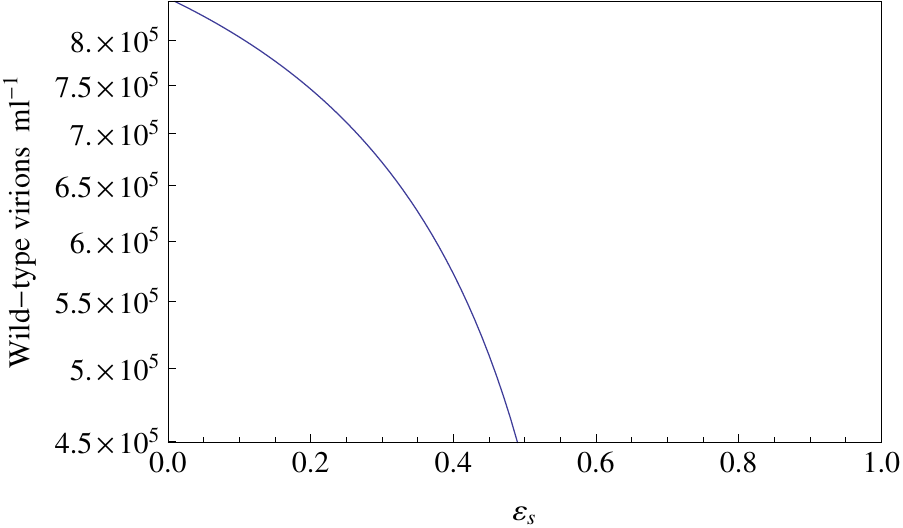}}  \quad 
\subfigure[Total virus steady state]{\includegraphics[height=.25\textwidth]{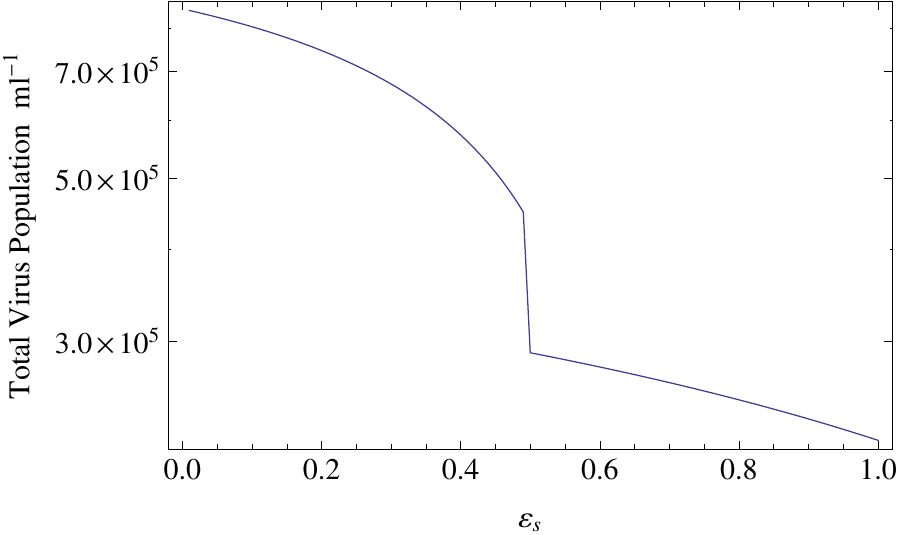}}
\caption{\footnotesize Dynamics of steady states for wild type and drug resistant virions with respect to drug efficacy, assuming $\alpha = 0.2$.}
\label{ViralLoad2}
\end{figure}

\section{Conclusions}
Upon formulating a new in-host model of HIV dynamics that incorporates latent infection and viral mutation, the local asymptotic behavior of the model was characterized and validated using computational means.  The effects of RTIs were elucidated under differing drug resistance levels for the mutated strain.  In particular, our analysis determined that a sufficiently high resistance level arising from viral mutation could render a persistent infection untreatable.  This implies that RTIs alone are not strong enough to counteract the emergence of drug resistant strains. Additionally, as HIV possesses a strong propensity to mutate, due to common errors in reverse transcription, the effects of viral mutation are significant within the dynamics of the virus.

\end{document}